Defining Domain-Independent Discovery Informatics*

William W. Agresti
Carey Business School
Johns Hopkins University
agresti@jhu.edu

ABSTRACT

This paper presents a personal account of the early legacy of discovery informatics, especially surrounding the first published definition of domain-independent DI. The immediate consequences of this DI definition and description are highlighted by examining citations in various reference sources, social media, educational and research programs, and the literature on the fourth paradigm of the scientific method. Indicative of DI's cross-cutting foundational and domain-independent aspects, this early definition and description are shown to be cited in research papers from multiple domains. Observations are offered on the state of discovery informatics, concluding that the timeless quest for knowledge and the relentless advances in informatics will propel DI to retain its appeal as a highly apt descriptor for research and practice activities that are inherent in our human nature.

1. INTRODUCTION

As activity under the heading of discovery informatics continues to evolve, I am prompted to summarize work that seemed then, and still to this day, to be the first definition and use of the term in its unrestricted form. While these reflections are necessarily personal, I hope they contribute to recording the emergence of this field of research and practice.

My defining of DI was foreshadowed by research and professional practice in empirical studies with large datasets (e.g., [7,8,9,14]). For Computer Sciences Corporation and MITRE Corporation, I supported principally NASA, defense, and national security initiatives. It was the big data and analytics of its time. Just prior to transitioning from industry to academia in 2000, my colleagues and I were focused on what an organization can know about itself from data that it already possesses (now, perhaps, an element of "business intelligence"). So it was knowledge discovery from data, but here it was helping an organization capture and capitalize on its knowledge, such as the ways that communication patterns can inform a company about which employees are its resident experts. We were building KM platforms to facilitate expertise location, collaboration, semantic search, content structuring, and explicit knowledge capture. [2]

---

When I started at Johns Hopkins University in 2000, my office was at its suburban Washington DC campus in the Shady Grove Life Sciences Center along I-270 in Montgomery County, Maryland. It was an exciting time and place with all of the activity to sequence the human genome. We were in the midst of bioscience companies and organizations, including many that I visited, such as Celera Genomics, Human Genome Sciences, and The Institute for Genomic Research (TIGR). Within this environment I became aware of the term "drug discovery informatics" to characterize the use of platforms of networked hardware and software for drug discovery. The term seemed like an apt label. But my background and experience (including getting further visibility into the research community as an NSF/CISE program director in the late 1990's) led me to believe that this condition (abundance of data and at increasingly finer granularity) was not exclusive to this domain. Granted, the dramatic human genome success had heightened expectations for what may be possible in bioinformatics, but the general circumstance was something I had seen and worked on in the domains of space science, defense, national security, and organizational knowledge. So, in 2001, I began drafting an essay on discovery informatics to make a case for recognizing the fundamental and broadly applicable advances that will be possible with a coherent approach to data-driven knowledge discovery. Seeking to reach a broad readership in an outlet with a large circulation and part of a prominent and searchable digital library, I published it in the *Communications of the ACM*, appearing in August 2003. [3]

## 2. DI IN COMPUTING AND THE SCIENTIFIC METHOD

Wanting to highlight the foundational nature of data-driven discovery in the CACM essay, I compared the state of DI to the evolution of the artificial intelligence field finding its place in the computing body of knowledge, as represented by the ACM Classification System. An early version of that organizing structure had AI as an application, one of many, for computing. The next version recognized the inadequacy of that positioning, and instead placed AI as a cross-cutting "function" in the taxonomy: there can be hardware, software, tools, methods, processes, etc. in AI research and practice. My claim in [3] was that DI needed to be similarly regarded because, like AI, it addresses fundamental questions (what are paths to new knowledge?), engages all aspects of computing, and is inherently cross-cutting. As hypothesis-free exploration of data, DI is a worthy approach to discovery.

Discovery informatics is often discussed in connection with paradigms for scientific investigation (see appendix). Some view DI as part of computational science, forming an emerging third model (after theory and experiment) in the scientific method. [26]. Others have cited my essay on DI when they describe and give a name to a distinct fourth paradigm of scientific research:
> "The greatest scientific research challenge with these massive distributed data collections is consequently extracting all of the rich information and knowledge content contained therein, thus requiring new approaches to scientific research. This emerging data-intensive and data-oriented approach to scientific research is sometimes called discovery informatics or X-informatics (where X can be any science, such as bio, geo, astro, chem, eco, or anything; Agresti 2003; Gray [and Szalay] 2003; Borne 2010). This data-oriented approach to science is now recognized by some (e.g., Mahootian and Eastman 2009; Hey et al. 2009) as the fourth paradigm of research, following (historically) experiment/observation, modeling/analysis, and computational science." [13]

When the passage above refers to my definition and description of DI and Jim Gray's X-informatics as two names for this fourth paradigm, it seems to me that, notionally, if we integrate over X, we get discovery informatics.

3. DEFINING AND DESCRIBING DI

With the publication of [3], I received many and diverse responses: notes from people agreeing with the idea and from companies that were dealing with massive data and wanting to discuss strategies to cope. As I began to see citations and explicit references, I found the article also directly cited in the creation of university laboratories, college courses, and at least one entire degree program. I started giving lectures (e.g., [5]) as an ambassador for DI, heralding this novel way of knowing, introducing the subject and the concepts to different audiences.

When I drafted the essay, by titling it Discovery Informatics, I believed that it would be a new term for readers. And nothing that I have read or heard since 2001 has led me to think otherwise. The previously mentioned and clearly more narrow and domain-specific "drug discovery informatics" would have been the only other similar term that readers may have experienced at that time, but, as discussed, likely only if they were in certain roles within bioscience and pharmaceutical fields. So, in consideration of the novelty of the title, I wanted to provide a definition. I had several objectives in mind for what I thought would be appropriate. I wanted the definition to convey what I considered the universality of the subject, its distinctiveness from query-driven discovery, and its applicability across domains and fields of study. Based on my experience with empirical investigations, I wanted to express a note of caution for practitioners not to claim cause-effect relationships at the first sign of revealed patterns in the data. I envisioned DI as "... the mode of scientific inquiry that produces 'starting points' for inductive and deductive reasoning." [30] I wanted the definition to give prominence to the need for validation of any patterns and relationships that emerge from the analytics. As a new way of knowing, data-driven discovery processes must recognize roles for verification and validation methods to establish knowledge claims. So, the resulting definition was, "Discovery Informatics is the study and practice of employing the full spectrum of computing and analytical science and technology to the singular pursuit of discovering new information by identifying and validating patterns in data." [3]

In 2005 and 2008 I continued to elaborate on the DI concept and methodology when I authored entries in the first and second editions of the *Encyclopedia of Data Warehousing and Mining*. [4, 6] Here I contrasted hypothesis-free DI with traditional hypothesis-guided discovery. I sought to explain DI as spanning domains as well as technologies, the latter broadly considered to include hardware-software processing platforms, storage architectures, data curation, visualization, semantic analysis, rule induction, machine learning, genetic algorithms, evolutionary computation, and instance-based learning.

4. EARLY DOMAIN-SPANNING DI ACTIVITIES

Before [3] appeared in print, Jay Liebowitz joined our faculty. As an expert and widely acknowledged thought leader in AI and KM, he was immediately a valued colleague and collaborator. We wanted to explore the breadth of this new field of DI. Based on our experiences, we believed there was outstanding data analytics work going on within domains, but not a lot of sharing across domains. So, a potentially useful way to help to advance this new field would be to draw leading experts out of their domains and have them share their DI approaches, methods, tools, and experiences. Our hope was that the exposure across domains would help start collaborations and informal relationships that would assist DI researchers and analysts as they returned to their domain-oriented investigations and as they may be encouraged to take on new knowledge-discovery challenges at domain boundaries. Through Jay's leadership, we launched a series of "Visionary Lectures in Discovery Informatics."

The lecture series brought in outstanding speakers in 2003-04 and had hundreds of people attend. The list of speakers and their topics demonstrated our commitment to explore the universality of DI across domains. As an indication of the intellectually rich local environment at the Johns Hopkins Montgomery County campus, we literally went across the street in one direction to get Steven Salzberg from TIGR (and who is now back at JHU), and across the street in the other direction to NASD (now FINRA) to get Henry Goldberg, one of the outstanding AI Ph.D.s using DI/analytics to enhance the integrity of the financial industry.

The list of speakers (with their then-current affiliations) and topics were as follows:
- "Emerging Trends in Data and Applications Security: Secure Semantic Web, Secure Knowledge Management, and Discovery Informatics," Dr. Bhavani Thuraisingham, Program Director, Data and Security Applications, National Science Foundation
- "The Role of Discovery Informatics in Financial Regulation and Surveillance," Dr. Henry Goldberg, Assistant Director, Knowledge and Data Discovery, Market Regulation, National Association of Securities Dealers (NASD)
- "Basic Principles and Methods for Developing, Validating, and Deploying Predictive Scores in the Business Environment Using Data Mining Tools," Hans Aigner, CEO, ASC Database Marketing; President, DataLab USA
- "Genome Sequencing and its Application to Microbial Forensics and Biodefense," Dr. Steven Salzberg, Senior Director of Bioinformatics, The Institute for Genomic Research (TIGR)
- "Discovery Informatics in Education," Dr. Ramon Barquin, President, Barquin International
- "Connecting the Dots: Graph-based Discovery Informatics for Learning Patterns of Asymmetric Threats," Dr. Larry Holder, Associate Professor, Department of Computer Science and Engineering, University of Texas at Arlington

- "Knowledge Discovery and Data Mining Based on Hierarchical Segmentation of Image Data," Dr. James C. Tilton, Applied Information Sciences Branch, NASA Goddard Space Flight Center

We continued our development of DI in education and research. We offered the first regular graduate course in Discovery Informatics at JHU in Fall 2003, and proposed a new Center (later, a Laboratory) for Discovery Informatics, described as …

> *"… an innovative approach to harness the vast capabilities of information technology and bring them to bear on data-driven discovery of knowledge that advances science, technology, organizations, and society. A need exists for a multidisciplinary approach for performing knowledge discovery across these domains to help create, capture, organize, share, and manage knowledge in a synergistic way. The center will bring a unifying presence to the emerging discipline of Discovery Informatics (DI), the study of informatics in support of discovery. DI pursues advances in information technology (algorithms, databases, processes, practices, and systems) that contribute to the discovery of knowledge. The rationale for the Center is that the discovery of knowledge can be advanced by exploring synergies and leveraging approaches and technologies that are now associated with a variety of diverse disciplines."*

The establishment of the laboratory was covered in the local press (e.g., [1]), giving visibility to the new DI term. Foundation funding for the lab enabled the support of graduate students to conduct research, leading to several publications (e.g., [12, 38]). Through funding by the GEICO Philanthropic Foundation, several students were recognized by a competitive award as GEICO Scholars in Discovery Informatics. As faculty members we had the satisfaction of seeing our graduate students being highly sought after by top employers.

5. EARLY LEGACY OF THE DI DEFINITION FROM [3]

Some DI definitions retain the original domain orientation to the biosciences. For example, even today, the online yourdictionary and wiktionary define DI as "the computing tools and methods used in the drug discovery process, such as software that deals with chemical and biological structures." [20]

In contrast, I wrote about DI as domain-independent: "what distinguishes discovery informatics is that it brings coherence across dimensions of technologies and domains to focus on discovery." [6]

Citations to the domain-independent definition of DI from [3] are found in reference materials and social media, books on data and informatics, and research papers on science methodology. It is noteworthy in support of the claimed breadth and foundational aspects of DI that this definition is also cited across domains such as astronomy, physiology, healthcare, and the life sciences.

5.1 Reference Materials and Social Media

- The DI definition from [3] is the one used in 2006 in the *Dictionary of Information Science and Technology* [19]:
    "Discovery Informatics:
    1. Knowledge explored in databases with the form of association, classification, regression, summarization/generalization, and clustering (Wu & Lee, 2005) 2. The study and practice […] in data. (Agresti, 2005)." (The citation is to [4], which includes the same definition from [3]).

- The *Encyclopedia of Data Warehousing and Mining* [4,6] includes an entry for DI that I was invited to write, so it is the definition from [3].

- In the popular professional social media site, LinkedIn, the group, International Society for Discovery Informatics, has 761 members (as of 2015). LinkedIn's description of the subject of the group mirrors what is described above regarding first an orientation to drug discovery and then its generalization in [3]:
    "The term 'Discovery Informatics' was first coined by Claus et al. (2002) for drug discovery process, and further generalized by Agresti (2003), as follows: "Discovery Informatics is the study and practice […] patterns in data." [29]

5.2   Books on Data, Informatics, and Knowledge Discovery

- From *Big Data Fundamentals: Concepts, Drivers, and Techniques*:

    "In 2003, William Agresti recognized the shift toward computational approaches and argued for the creation of a new computational discipline named Discovery Informatics. Agresti's view of this field was one that embraced composition. In other words, he believed that discovery informatics was a synthesis of the following fields: pattern recognition (data mining); artificial intelligence (machine learning); document and text processing (semantic processing); database management and information storage and retrieval. Agresti's insight into the importance and breadth of computational approaches to data analysis was forward-thinking at the time, and his perspective on the matter has only been reinforced by the passage of time and emergence of data science as a discipline." [21, p. 182]

- In *The Book of Informatics*, Gammack, et al. write:
    "With very large tables of data, computers would do this type of analysis, or using more advanced techniques of discovery informatics, defined by Agresti as 'the study and practice […] in data.' If you mine the nuggets from Agresti's definition it reduces to 'using pattern-finding technology to discover new information which would then be validated and related to existing human understanding. New discoveries can emerge from this, such as …" [22, p. 108]

- In *Knowledge Development Innovation*, Miller writes,
    "An earlier reference was made to William W. Agresti in the data section. He is also recommending a new methodology that he describes as 'discovery

informatics.' He defines the term as the study and practice […] patterns in data. … Professor Agresti describes discovery informatics as a methodology that should be recognized and included in the ACM (Association for Computing Machinery) classification system. (Agresti 2003)" [32, p. 35]

5.3  DI and Science Methodology

Titles of various research papers and presentations on science methodology suggest the range of instances in which the DI description and definition from [3] have been cited:
- "Is Discovery Informatics a Methodology or an Ontology?" Keynote Speech, International Arab Conference on Information Technology, Jordan:
  "In 2003 Professor Agresti titled his article in the ACM (Association for Computer Machinery) with a new term 'Discovery Informatics' [1]. He describes the Discovery Informatics in the Encyclopedia of Data Warehousing and Mining [2] as an emerging methodology that promotes a crosscutting and integrative view. Discovery Informatics looks across both technologies and application domains to identify and organize the techniques, tools, and models that improve data-driven discovery." [23]

- "Discovery Informatics: An IE perspective":
  "I read the article, and was impressed with Agresti's identification of discovery informatics. It is logical … that the field Agresti calls discovery informatics will become increasingly significant with the years." [10]

- "Mopping up the flood of data with web services" [39]

- "Of mice and mentors: Developing cyber-infrastructure to support transdisciplinary scientific collaboration" [25]

- "Informatics Creativity: A Role for Abductive Reasoning?" [36]

- "Grid-Enabled Measures: Using Science 2.0 to Standardize Measures and Share Data" [33]

5.4  Domains of Research and Practice

5.4.1 Astronomy

The DI definition and description from [3] figured prominently in a position paper prepared and endorsed by a team of 91 astronomers and information scientists for the Astro2010 Decadal Survey:
  "A Vision for Astroinformatics – The New Paradigm for Data-Intensive Astronomy. Astroinformatics is Discovery Informatics for astronomy. We believe that it can and should become a standalone research discipline. Agresti [33] defined *Discovery Informatics* as: *"the study and practice […] patterns in data."*

The late Jim Gray (of Microsoft Research) championed the development of this fourth leg of science (data-intensive science), which he called X-Informatics (in KDD-2003), to accompany theory, computing, and experiment (observation). For our discipline, X is "astronomy". An informatics paradigm is needed within any data-intensive scientific discipline to make the best use of its rich data collections for scientific discovery. Discovery Informatics thereby activates data integration and fusion across multiple heterogeneous data collections to enable scientific knowledge discovery and decision support. . . . Discovery Informatics is therefore a key enabler for new science discovery in large databases, through the application of common data integration, browse, and discovery tools within a discipline …" [34, p. 5]

5.4.2 Physiology

In a paper in the *Journal of Physiology*, Silva discusses methods and practices in scientific research: "Central to the genesis and growth of any scientific discipline is the development of tools for the study and objective evaluation of relevant phenomena and ideas." [37, p. 206] When he begins introducing experimental frameworks and tools, his first such framework is DI, citing the definition from [3]:
> "The term 'discovery informatics' was first defined by William W. Agresti in 2003, as follows: 'Discovery Informatics is the study and practice […] patterns in data.' (Agresti, 2003) … In its most idealized conception, discovery informatics … uses the power and relentlessness of computers to find unrecognized connections … Science often progresses by connecting sets of information that evolved independently (i.e., bridging areas or even fields of research) … the astronomical number of all possible connections is such that computers could make a real contribution with this association process." [37]

5.4.3 Healthcare

In a chapter on "Public Health Informatics," B. W. Hesse states that, "Whereas the emphasis during the previous decade was on new data collection and storage , the emphasis during this new era will be on 'connecting the dots' within the data already collected, an emphasis some have referred to as an era of 'Discovery Informatics' (Agresti, 2003)." [24]

5.4.4 Life Sciences

From a blog on Life Sciences Discovery Informatics:
> "There is a nascent field called "Discovery Informatics" which is devoted to applying computer science to advance discovery across all scientific disciplines. The field is so nascent, in fact, that Wikipedia has nothing to say about it. The best definition I could find is this one from William W. Agresti [1]: 'Discovery Informatics is the study and practice […] patterns in data.'" [28]

6. DI IN EDUCATIONAL AND RESEARCH PROGRAMS

The definition and essay in [3] were cited in the creation of an undergraduate degree program, a BS in Discovery Informatics, in Fall of 2005 at the College of Charleston:

- "Origin of the Name and Concept. William Agresti, in an August 2003 article [2], coined the term "discovery informatics" to refer to "the study and practice of employing the full spectrum of computing […] in data." … The Agresti article appeared while our committee was working on this new undergraduate curriculum. The similarity between what Agresti defined as 'discovery informatics' and the curriculum that was already in development inspired us to choose Discovery Informatics (DI) as the original name for the new degree program." [11]

- From the program description on the web site [16]:
  "The term 'Discovery Informatics' was defined by William W. Agresti in 2003, as follows: 'Discovery Informatics is the study and practice […] patterns in data'."

- Media reported on the new College of Charleston degree program in DI [17]:
  - "Discovery Informatics wasn't even a term until 2003"
  - "Discovery Informatics, an emerging science that is on track to become the newest undergraduate major at the College of Charleston -- and the first program of its kind in the United States."
  - "…at the field's leading edge, Discovery Informatics attempts to bridge the gap between the limits of human intelligence and the mind-boggling ocean of information that human intelligence created."
  - "'Yes, it is very clever, and, in fact, Clemson and Carolina are very envious that they didn't think of it first,' said Noonan, dean of the college's School of Science and Mathematics. 'And (President Lee Higdon) was intrigued about the opportunity to bring the college some distinctiveness.'"

The description and definition of DI in [3] were cited in the creation of various concentrations, specializations, courses, and research centers on the subject, in the US and internationally:
- Rochester Institute of Technology has a concentration in Discovery Informatics in its MS in Information Sciences and Technology program (http://it.rit.edu/).
- College of Barcelona, School of Informatics: specialty in Data Mining and Business Intelligence (http://www.fib.upc.edu/fib/centre/govern/organs-colegiats/actes/mainColumnParagraphs/0111117/document2/Justificacio_nova_especialitat_MIRI.pdf)
- National Institute of Oceanography, India: Course No: CBB (NISCAIR)-204 Discovery & Translational Bioinformatics (www.nio.org)
- Center for Discovery Informatics at Rochester Institute of Technology. [35]: "The mission of the CDI is to conduct high quality research in the application of computational techniques to the management and understanding of data-intensive systems. … Why Discovery Informatics? As William Agresti points out in his recent article in Communications of the ACM, the world is being flooded in data. Extracting knowledge, information, and relationships from this data is one of the greatest challenges faced by the scientists in the twenty-first century."

## 7. OBSERVATIONS ON THE STATE OF DI

There is a steady stream of intellectual activity and progress in DI. Research centers and laboratories featuring DI in their names include:
- Rutgers Discovery Informatics Institute
- Penn State Center for Big Data Analytics and Discovery Informatics
- Indiana University School of Informatics, Indianapolis, and Purdue University School of Science Department of Computer & Information Science, Indianapolis: Discovery Informatics and Computing Laboratory

In other cases, laboratories that had DI in their title have transitioned to having changed it to some variations of data science or analytics. There are additional examples of DI figuring prominently in the work of other research centers, notably Purdue's Center for Catalyst Design and the University of Washington-Tacoma's Center for Data Science.

DI continues to be the subject of workshops and conferences, such as --
- Web Observatories for Discovery Informatics, July 18, 2012, Microsoft, Redmond, WA
- Carnegie Mellon's CMUSV Symposium on Cognitive Systems and Discovery Informatics, June 21-22, 2013
- AAAI Fall Symposium: Discovery Informatics: AI Takes a Science-Centered View on Big Data, November 15-17, 2013, Arlington, VA
- NSF Workshop on Discovery Informatics, February 2-3, 2014, Arlington, VA

The discovery informatics initiative, http://www.discoveryinformaticsinitiative.org, is a very useful hub for DI activities, including publicizing conferences and workshops on DI-related research and practice.

There are observations on the evolution of the field. For example, consider the characterizations of DI from organizers of the Microsoft and Carnegie Mellon workshops noted above:
- "Discovery informatics is emerging as a three-area discipline with themes around computational support of discovery processes (such as knowledge bases, provenance standards, and visualizations), the study of the interplay between data and models (such as tradeoff between expressiveness and scalability), and the use of social computing for discovery (such as management of human contributions)." [31]
- "Scientific research is one of the most complex activities in which humans engage, and this enterprise is becoming ever more complex, dealing with more data and with more sophisticated models than ever before. This daunting complexity has led to the emerging field of *discovery informatics*, which aims to: (a) understand the representations and processes that underlie scientific research; (b) develop computational artifacts that embody these understandings;

and (c) apply these systems to specific scientific problems to assist in obtaining new research results." [15]

There are noticeable trends in the evolution of DI both as a field of research and practice, and as a term. In many cases, data science and analytics have emerged as preferred terms in the titles of degree programs, research centers, and job titles – and sometimes as replacements for DI (e.g., The College of Charleston changed the name of its degree program from DI to data science in 2012). [11] I believe these other terms are overlapping but do not convey the same ambition as DI. This naming business runs a natural course, with various drivers and influences, such as what works best to represent the actual content of the academic studies or research investigations or job descriptions – or what resonates better with prospective customers, potential collaborators, funding sources, and the public.

I find it heartening that DI retains such steady recognition, through its continued use in workshops and research initiatives, as the appropriate term for so much vitally important work. DI conveys a clear message as to the intention of the studies and research: informatics in support of knowledge discovery. Data science, data analytics, big data, and predictive analytics, while possessing their own attractive qualities as rubrics are not *prima facie* about knowledge discovery.

KDD is not the same as data mining, yet draws upon it. DI is not the same as data science and analytics, yet will benefit from advances in those fields. Both KDD and DI are purpose-carrying titles, pursuing the never-ending quest for discovery.

8. CONCLUSIONS

I and others have emphasized the foundational character of DI in connection with paradigms for scientific methods – noted above, with DI as part of the third paradigm, or, more convincingly, defining the fourth paradigm all by itself. As impressive as it may be for DI to be considered one of the four scientific paradigms, I also have stressed in [3] and reinforce here, that there is even a more fundamentally inherent nature of DI for creative discovery as constituting the automated counterpart to the human subconscious, unconstrained by intentionality. As we consider creativity arising in humans from both conscious and subconscious intellectual activity, before DI there was only intentional automated support for knowledge acquisition. And yet, "… most breakthroughs are based on linking information that usually is not thought of as related." [18] Discovery informatics is the heretofore missing complement -- a kind of automated subconscious for data-induced knowledge discovery --- where "rationality cannot censor the connection." [18]

## Appendix

## Discovery Informatics and the Fourth Paradigm

I first saw it at the Orsay Museum in Paris. Over six feet in height, "Nature Unveiling Herself to Science" is an imposing and dramatic allegorical sculpture. This single object melds two defining features of Johns Hopkins scholarship: as a work of art, creativity; as an allegory, discovery. And, as for this unveiling part, how does that happen? What are the methods of Science to uncover Nature?

From the first glimmer of intelligent life on the planet until about 50 years ago, if we enumerated the scientific paradigms, we may only reach the grand total of two. As noted in the mid-1980s, "Science is undergoing a structural transition from two broad methodologies to three – namely from experimental and theoretical science to include an additional category of computational and information science. A comparable example of such change occurred with the development of systematic experimental science at the time of Galileo." [27]

In 1987 Nobel laureate Ken Wilson characterized computational science as the third paradigm of Science, as simulation increasingly became the only viable methodology to accomplish some critical unveilings of Nature. Concurrently, data-driven discovery was beginning to emerge as a distinctive methodology on its own, constituting the fourth paradigm. This contrast to hypothesis-driven methods brings to mind another Johns Hopkins connection with C. S. Peirce's 1899 admonition, "Do not block the path of inquiry." Discovery informatics offers a new path to inquiry.